\documentclass[aip,amsmath,amssymb,reprint,superscriptaddress,floatfix]{revtex4-2}

\usepackage{graphicx}% Include figure files
\usepackage{floatrow}
\usepackage[caption=false,font={large}]{subfig}
\usepackage{bm}% bold math

\usepackage{amsmath}
\usepackage{amssymb}
\usepackage{color}
\usepackage{times}
\usepackage{float}
\usepackage[sort&compress]{natbib}

\usepackage[colorinlistoftodos,prependcaption,textsize=small]{todonotes}

\usepackage{hyperref}

\hypersetup{
  colorlinks=true,
citecolor=black,
linkcolor=black,
urlcolor=black
  }

\usepackage[T1]{fontenc}
\usepackage[utf8]{inputenc}

\begin{document}

\title{Reset Induced Multimodality in Unbounded Potential}

\author{Karol Capa{\l}a}
\email{capala@agh.edu.pl}

\affiliation{Institute of Computer Science, AGH University of Krakow, Kawiory 21, 30-059 Krak\'ow, Poland}

%Lines break automatically or can be forced with \\

\date{\today}% It is always \today, today,
             %  but any date may be explicitly specified

\begin{abstract}
Resetting, as a protocol that restarts the evolution of a system, can significantly influence stochastic dynamics. 
One notable effect is the emergence of stationary states in unbounded potentials, where such states would otherwise be absent without resetting. 
In this work, we explore unbounded potentials for which resetting not only induces stationary states but also leads to their multimodality, despite the repulsive nature of the potential. 
We present examples of potentials that, despite lacking local minima, can generate trimodal and pentamodal states, and we investigate how the modal structure of these states varies with noise intensity and resetting frequency. 
\end{abstract}

% \pacs{02.70.Tt,
%  05.10.Ln, %Monte Carlo methods statistical physics and nonlinear dynamics,
%  05.40.Fb, % Random walks and Levy flights
%  05.10.Gg, % Stochastic analysis methods (Fokker--Planck, Langevin, etc.)
%   02.50.-r, % Probability theory, stochastic processes, and statistics
%   }

%
\maketitle

\setlength{\tabcolsep}{0pt}

\textbf{
Stochastic processes provide a powerful framework for modeling diverse physical phenomena. 
One of the fundamental problems in stochastic process theory is the analysis of stationary states.
In systems driven by Gaussian white noise, stationary states typically mirror the underlying system symmetries. 
However, this correspondence breaks down for nonequilibrium scenarios. 
This work investigates non-equilibrium stationary states in unbounded potentials subject to Poissonian resetting. 
We establish that such potentials can sustain multimodal stationary states even in the absence of local minima. 
The results were obtained for three different potentials, demonstrating parameter ranges of resetting rate and noise intensity where modes emerge exclusively from resetting, depend solely on the potential's shape, and exhibit intermediate regions where both modal types coexist. 
These findings advance both the specific understanding of resetting phenomena and our fundamental knowledge of multimodal non-equilibrium stationary states.
}

%%%%%%%%%%%%%%%%%%%%%%%%%%%%%%%%%%%%%%%%%%%%%%%%%%%%%%%%%%
%
%
\section{Introduction}
Random walks are ubiquitous in the modeling and analysis of diverse systems. 
Typically, the environment is assumed to be in equilibrium, with fluctuations modeled as Gaussian white noise, leading to stationary states described by Boltzmann distributions~\cite{gardiner1983,risken1984}. 
In such equilibrium conditions, the symmetry of the potential is directly reflected in the stationary states, specifically, the number and location of the modes coincide with the potential’s minima. 
However, there is a need to explore nonequilibrium stationary states, where deviations from equilibrium assumptions can reveal novel phenomena and complex modal structures.

A potential source of nonequilibrium in a system is the presence of nonequilibrium noise, such as L\'evy noise. The emergence of these nonequilibrium stationary states requires more stringent conditions, in particular, the potential must be sufficiently steep~\cite{jespersen1999,dybiec2010d}. 
Conversely, L\'evy noise can induce multimodal stationary states even in single-well potential~\cite{chechkin2002,chechkin2003,chechkin2004,chechkin2006,chechkin2008introduction,capala2019multimodal,capala2019underdamped}. 
Notably, the curvature of the potential is a critical, though not exclusive, factor that determines the number and position of the modes~\cite{chechkin2002,capala2019multimodal}.

Similar behavior can also be observed for other types of noise that have the potential to drive the particle far from origin.
In particular, fractional Brownian noise imposes additional constraints on the formation of stationary states~\cite{guggenberger2022absence} and can lead to the emergence of multimodal states~\cite{guggenberger2021fractional}. 
Likewise, multimodal stationary states in a quartic potential have been observed in the Ornstein-Uhlenbeck process~\cite{dybiec2024multimodality}.

Another source of nonequilibrium behavior may arise from resetting, a protocol in which the system's dynamics are restarted from their initial state~\cite{evans2011diffusion,evans2020stochastic,gupta2022stochastic}.
Intervals between resets can be periodical (sharp resetting)~\cite{pal2017first} or follow diverse statistical distributions. 
While the most commonly studied case is Poissonian resetting~\cite{evans2011diffusion}, where resets occur at a constant rate, alternative approaches include power-law~\cite{nagar2016diffusion}, Weibull, log-normal, log-logistic, and Fr\'echet distributions~\cite{reuveni2016optimal,pal2017first}. 
Moreover, although resetting typically restores the system to its initial state, protocols are also considered in which the particle is only partially reset, meaning it is restored to a random point between its current and initial positions~\cite{dahlenburg2021stochastic,tal2022diffusion,bello2023time}, or even relocated to a completely random point~\cite{pal2024random}.
Beyond classical systems, the impact of resetting on dynamics has also been studied in the context of quantum random walks~\cite{mukherjee2018quantum,rose2018spectral,yin2023restart,yin2025restart}.

For the present study, it is particularly relevant that resetting can give rise to (nonequilibrium) stationary states~\cite{eule2016non}, what was explored in variety of scenarios, including free motion \cite{evans2011diffusion,nagar2016diffusion}, L\'evy flights \cite{stanislavsky2021optimal}, continuous time random walks \cite{mendez2021ctrw} or even motion in inverted potentials \cite{pal2015diffusion}. 
It can also influence stationary states in bounded potentials~\cite{pal2015diffusion,trajanovski2025generalised}.
It has even been shown that in systems driven by $\alpha$-stable noise, which exhibit bimodal states in the absence of resetting, an appropriate choice of the resetting rate can either induce trimodal states or completely eliminate multimodality~\cite{pogorzelec2023resetting}.

This study explores the possibility of multimodal stationary states induced by resetting in systems driven by Gaussian noise within a potential that features a barrier and then rapidly decays to zero at infinity. In the absence of resetting, such a system would not exhibit a stationary state, let alone a multimodal one. The paper is structured as follows: Section~\ref{sec:model} introduces the investigated model, Section~\ref{sec:results} presents the results, and the final Section~\ref{sec:summary} provides a summary and conclusions.

%%%%%%%%%%%%%%%%%%%%%%%%%%%%%%%%%%%%%%%%%%%%%%%%%%%%%%%%%%
%
%
% \clearpage
\section{Model\label{sec:model}}

The Langevin equation~\cite{gardiner1983} is given by:

\begin{equation}
    \frac{d x}{dt}= -V'(x) + \xi(t),
    \label{eq:langevin}
\end{equation}
where \(\xi(t)\) represents Gaussian white noise satisfying \(\langle \xi(t) \rangle = 0\) and \(\langle \xi(t) \xi(s) \rangle = \sigma^2 \delta(t-s)\). Here, \(V'(x)\) denotes the derivative of the potential \(V(x)\) with respect to the position \(x\).

The focus of the following research is on potentials that exhibit a global maximum at \(x = 0\) and then monotonically decrease to $0$ as \(x \to \pm \infty\). To analyze the conditions under which multimodal stationary states arise, two specific potentials were selected, which are expected to generate a trimodal stationary state, and one additional potential was chosen to verify the occurrence of higher modality, specifically a five-modal state.

The first potential under investigation is an arctangent-type potential, given by Eq.~\eqref{eq:arctan-potential}. This potential is the negative of the one used in the Ref.~\onlinecite{capala2020levy} and was selected because it serves as a model for a rectangular potential barrier:

\begin{equation}
\label{eq:arctan-potential}
V(x)=-\frac{h}{\pi} \arctan{\left( nx^2-n\right)}.
\end{equation}
The parameter \(h\) controls the height of the potential, while \(n\) determines its ``rectangularity''. In the limit \(n \to \infty\), the potential given by Eq.~\eqref{eq:arctan-potential} approaches a rectangular barrier. Even for relatively small values of \(n\), this potential is relatively flat around \(x = 0\), then drops sharply, and finally becomes nearly flat again. This behavior makes it a suitable candidate for studying the emergence of multimodal stationary states in systems governed by stochastic dynamics with resetting.

The second potential under consideration is the Gaussian potential defined in Eq.~\eqref{eq:gaussian-potential}:
\begin{equation}
V(x)= h e^{-x^2}.
    \label{eq:gaussian-potential}
\end{equation}
Unlike the arctangent potential, this Gaussian potential exhibits curvature at $x=0$. This structural difference may influence the distinct probability distribution behavior emerging from frequently reset trajectories.
Nevertheless, the Gaussian potential (Eq.~\eqref{eq:gaussian-potential}) shares key characteristics with the arctangent potential (Eq.~\eqref{eq:arctan-potential}) as both possess a global maximum at $x=0$ and decay monotonically to $0$ as $x \to \pm\infty$.

\begin{figure}[!h]
    \centering
    \includegraphics[width=0.9\linewidth]{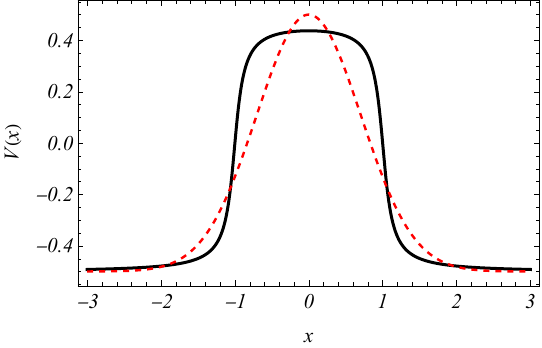}
    \caption{Arctangent potential barrier given by Eq.~\eqref{eq:arctan-potential} with $n=5$ (black solid line) and Gaussian potential barrier given by Eq~\eqref{eq:gaussian-potential} (red dashed line).} 
    \label{fig:potential}
\end{figure}

The examples of potentials described by Eqs.~\eqref{eq:arctan-potential} and \eqref{eq:gaussian-potential} are shown in Fig.~\ref{fig:potential}.

\begin{figure}[!h]
    \centering
    \includegraphics[width=0.9\linewidth]{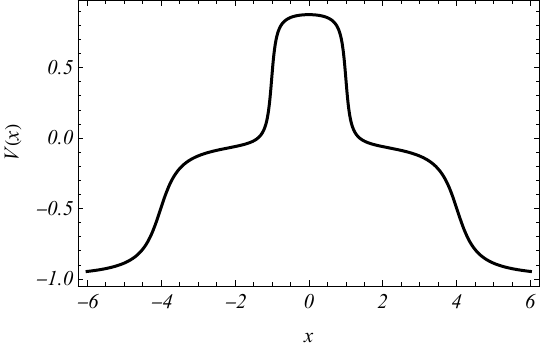}
    \caption{Arctan potential given used to explore five-modal stationarystates, given by Eq~\eqref{eq:multi-arctan-potential}}
    \label{fig:potential-multimodal}
\end{figure}

Finally, to investigate the emergence of higher modalities, a modified arctangent-type potential given by Eq.~\eqref{eq:multi-arctan-potential} was used.
\begin{equation}
\label{eq:multi-arctan-potential}
V(x)=-\frac{h}{\pi} \left(\arctan{\left( nx^2-n\right)} + \arctan{\left( n\left(\frac{x}{4}\right)^2-n\right)} \right) .
\end{equation}
The potential described by Eq.~\eqref{eq:multi-arctan-potential} forms a two-level "pyramid" composed of two arctangent potentials with different widths combined together. 
This constructed potential exhibits behavior similar to the potential given by Eq.~\eqref{eq:arctan-potential} in the vicinity of $x = 0$ and in the asymptotic regions ($x \to \pm\infty$), while exhibiting additional ``shelves'' with slight slopes that may enable the formation of additional modes in the stationary state.
The shape of the potential is shown in Fig.~\ref{fig:potential-multimodal}.

The choice of potentials can be further motivated by insights from multimodal stationary states induced by L\'{e}vy flights, where the occurrence of multimodal states exhibits an empirical relationship with the curvature of the potential~\cite{chechkin2003,chechkin2004,capala2019multimodal}. 
For a plane curve described by \(V(x)\), the curvature is given by:
\begin{equation}
\kappa(x)=\frac{V''(x)}{[1+V'(x)^2]^{3/2}}.
    \label{eq:curvature}
\end{equation}
The potentials defined by Eqs.~\eqref{eq:arctan-potential} and \eqref{eq:gaussian-potential} feature two curvature maxima located away from the reset point, while the potential in Eq.~\eqref{eq:multi-arctan-potential} exhibits four curvature maxima.

The system's Langevin dynamics (Eq.~\eqref{eq:langevin}) was extended with stochastic resetting via a Poisson process (rate $r > 0$), where the time intervals \( \tau \) between consecutive resets follow an exponential distribution:
\begin{equation}
\phi(\tau) = r e^{-r\tau}.    
\end{equation}
This leads to resetting events with no temporal correlations, where the probability of resetting remains constant in time, characterized by an average reset interval of \( \langle \tau \rangle = 1/r \).
Each reset event instantaneously returns the system to its initial position \( x_0 \equiv x(0) = 0 \), completely independent of previous evolution history.

A deeper insight into the process can be gained through an examination of the Fokker-Planck-Smoluchowski equation. 
For the steady-state scenario with resetting to $x=0$, the equation is given by~\cite{pal2015diffusion}
\begin{equation}
0 =  \frac{d}{d x}\left(V^{\prime}(x) P(x)\right) +\sigma^2 \frac{d^2 P(x)}{d x^2} - r P(x) + r \delta\left(x-x_0\right)
\label{eq:fps}
\end{equation}
where $P(x)$ denotes the stationary probability distribution.
An analysis of Eq.~\eqref{eq:fps} reveals that among the three parameters $\sigma$, $r$, and $h$, any two can be expressed as functions of the third. 
Consequently, in the subsequent discussion, we disregard the dependence on the potential barrier height to streamline the analysis.

The numerical results were obtained by integrating the Langevin equation~(\ref{eq:langevin}) using the Euler-Maruyama method \cite{higham2001algorithmic,mannella2002} with a time step of $\Delta t = 10^{-3}$. 
The results were averaged over $N = 10^6$ independent realizations, and from the generated trajectories, stationary states were analyzed and constructed.

%%%%%%%%%%%%%%%%%%%%%%%%%%%%%%%%%%%%%%%%%%%%%%%%%%%%%%%%%%%%%%%%%%%%%%%%
%
%
% \clearpage
\section{Results\label{sec:results}}
We begin the analysis with the non-equilibrium stationary states genereted by arctangent potential given by Eq.~\eqref{eq:arctan-potential}. 
The results presented in this section correspond to the case of $n = 5$; however, a preliminary analysis for other values of $n$ revealed qualitatively similar behavior.

\begin{figure}[!h]
    \centering
    \includegraphics[width=0.9\linewidth]{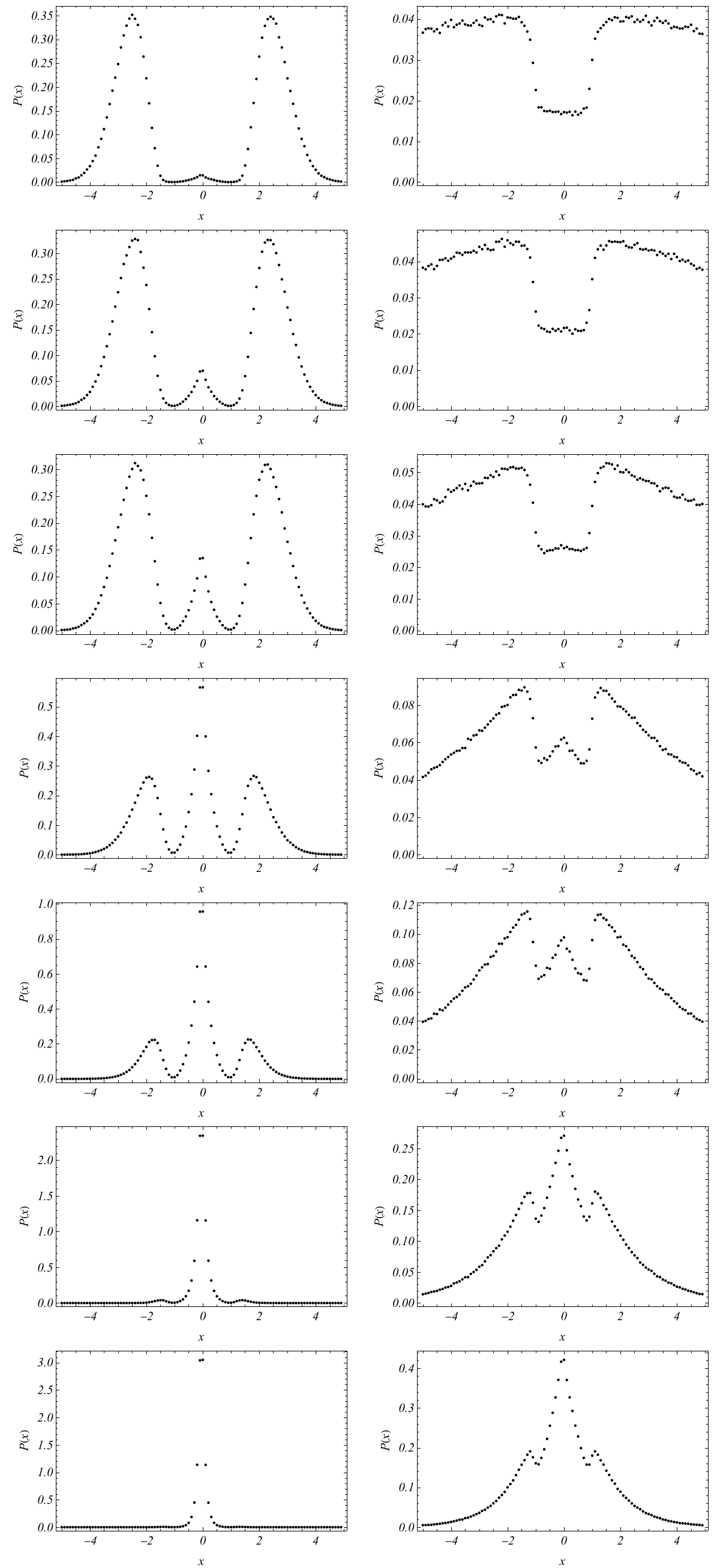}
    \caption{Stationary states for arctangent potential given by Eq.~\eqref{eq:arctan-potential} for noise intensity $\sigma=0.1$ (left column) and $\sigma=1$ (right column) for various resetting rates, $r=0.001,0.005,0.01,0.05,0.1,0.5,1$ from top to bottom.}
    \label{fig:arctan-states}
\end{figure}

The results for two representative values of $\sigma$, namely $\sigma = 0.1$ and $\sigma = 1$, are presented in Fig.~\ref{fig:arctan-states}. 
For $\sigma = 0.1$, when $r$ is small, two distinct outer maxima are observed, accompanied by a small peak at $x = 0$. 
As $r$ increases, the amplitude of the outer peaks decreases, while the inner peak becomes more pronounced. 
At sufficiently large values of $r$, the outer peaks vanish, and the system transitions to a unimodal distribution.
In the case of $\sigma = 1$, for small values of $r$, only the outer maxima are visible, i.e. stationary distribution is bimodal, and the distribution slowly decays as $x \to \pm \infty$. 
As $r$ increases, the outer peaks become narrower as more of the probability mass is located closer to origins. 
Around $r \sim 0.01$, an internal peak emerges at $x = 0$, and as $r$ continues to grow, the central peak strengthens while the outer peaks diminish. 
For values of $r$ larger than those shown in Fig.~\ref{fig:arctan-states}, the system generates a unimodal state, with only one mode located at the reset position.

\begin{figure}[!h]
    \centering
    \includegraphics[width=0.9\linewidth]{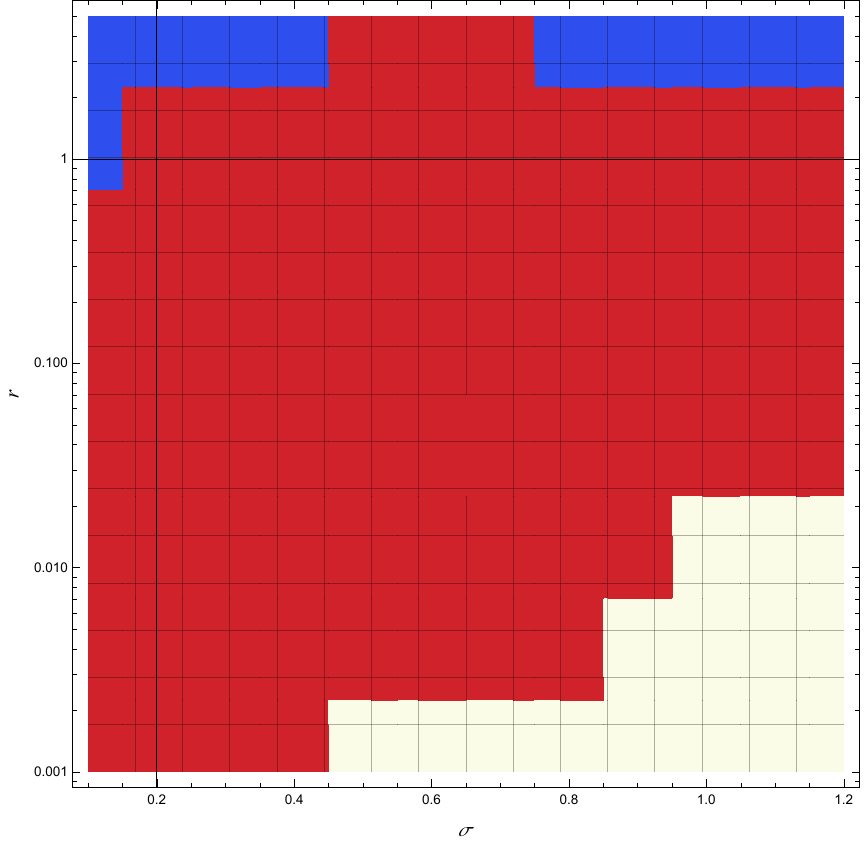}
    \caption{The figure depicts, in distinct colors, the parameter sets corresponding to stationary states of the arctangent potential, given by Eq.~\eqref{eq:arctan-potential}, with mode numbers three (red), two (white), and one (blue).}
    \label{fig:arctan-modes}
\end{figure}

Fig.~\ref{fig:arctan-modes} displays the number of modes in the stationary state distribution across parameter space for the arctangent potential.
The parameter space reveals three characteristic regimes.
Large resetting rates $r$ produce unimodal states, visible as the blue region, where frequent resetting dominates the dynamics.
When the noise intensity $\sigma$ dominates the resetting rate $r$ ($\sigma \gg r$), the system forms bimodal stationary states, corresponding to the white region in Fig.~\ref{fig:arctan-modes}.
The intermediate (red) region displays trimodal states with three peaks, manifesting when resetting's localization competes with diffusion's ability to explore the potential's structure.

\begin{figure}[!h]
    \centering
    \includegraphics[width=0.9\linewidth]{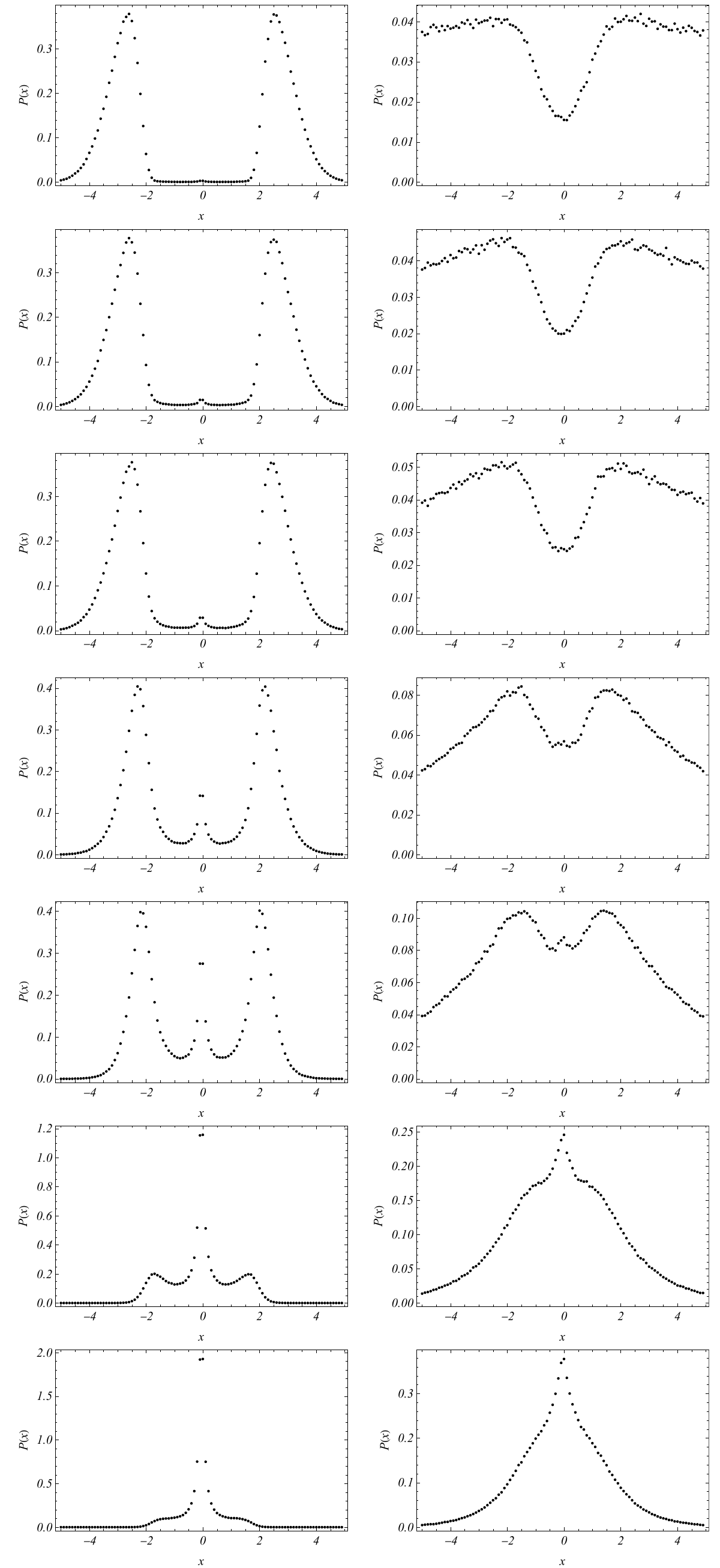}
    \caption{Stationary states for Gaussian potential given by Eq.~\eqref{eq:gaussian-potential} for noise intensity $\sigma=0.1$ (left column) and $\sigma=1$ (right column) for various resetting rates, $r=0.001,0.005,0.01,0.05,0.1,0.5,1$ from top to bottom.}
    \label{fig:gauss-states}
\end{figure}

\begin{figure}[!h]
    \centering
    \includegraphics[width=0.9\linewidth]{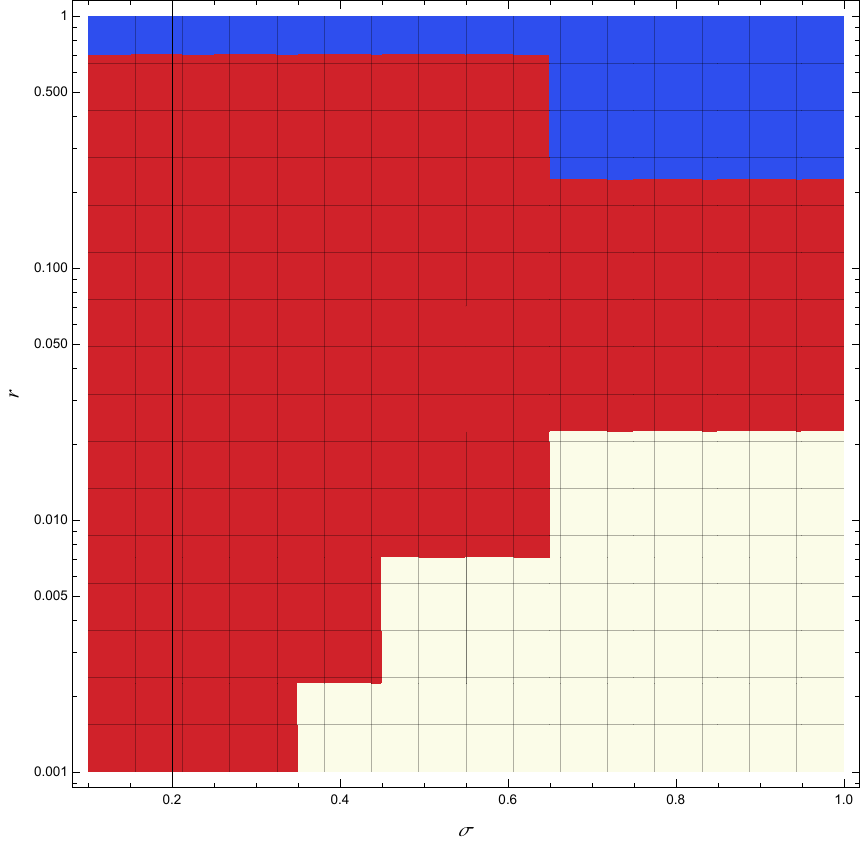}
    \caption{The figure depicts, in distinct colors, the parameter sets corresponding to stationary states of the Gaussian potential, given by Eq.~\eqref{eq:gaussian-potential}, with mode numbers three (red), two (white), and one (blue).}
    \label{fig:gauss-modes}
\end{figure}

The stationary states for the Gaussian potential, given by Eq.~\eqref{eq:gaussian-potential}, and their corresponding modal structures across parameter space are presented in Figs.~\ref{fig:gauss-states} and \ref{fig:gauss-modes}, respectively. 
While qualitatively similar to the arctangent potential results, the Gaussian potential displays three characteristic regimes: a unimodal region under frequent resetting, a bimodal region with weak resetting, and an intermediate trimodal region. 
Notably, the transition to bimodality occurs at smaller values of $\sigma$ and the system reaches unimodality at lower values of $\sigma$ compared to the arctangent case.
Clear differences are also evident in the stationary state distributions themselves (see Fig.~\ref{fig:gauss-states}).  
These are particularly pronounced for bimodal states, where the distributions reflect distinct potential shapes.  
While the arctangent potential produces distributions where the interpeak region mirrors the quadratic nature of the potential barrier, the Gaussian potential yields a U-shaped gap between peaks.  
This difference manifests in the transition to trimodal states: the local minimum near the origin first flattens before an additional mode becomes visible.  
Another distinction appears during the transition to unimodality. 
Here, unimodality emerges through the merging of inner and outer peaks, resulting in a sharp peak at the resetting point surrounded by a low but broad base.  
This base gradually narrows and eventually vanishes as $r$ increases.

The stationary states exhibit three distinct dynamical regimes governed by the competition between diffusion and resetting processes. 
In the resetting-dominated regime (large $r$), frequent resetting events localize particles near $x=0$, producing a unimodal distribution centered at the resetting point. 
The strong resetting effectively suppresses diffusive exploration of the potential landscape.

Conversely, in the diffusion-dominated regime (large $\sigma$), the stationary states resemble time-dependent probability distributions of reset-free evolution, effectively frozen at a particular time. 
Although resetting occurs too infrequently to confine particles spatially, it nevertheless maintains existence of non-equilibrium stationary states, resulting in bimodal distributions with peaks determined by the potential's shape.

The intermediate regime emerges from the balance between these competing effects. 
The stationary state displays both a central peak at $x=0$ (due to particles being frequently reset before escaping far from the origin) and outer peaks associated with potential-driven diffusion. 
The central peak reflects the spatial confinement imposed by resetting, while the outer peaks form in regions where the potential's slope becomes gentler. 

This spatial structure arises because the potential's gradient strongly repels particles toward $\pm\infty$, with the repulsive force intensifying in steeper regions. 
Consequently, particles accumulate in regions where the potential slope becomes gentler. 
In the absence of resetting, continued evolution in such potentials would inevitably lead to escape to infinity.
The resetting mechanism interrupts this escape by periodically returning particles to the potential maximum at $x=0$, creating a dynamical equilibrium where particles are continuously recycled into high-force regions while being prevented from distant exploration. 
This competition between potential-driven diffusion and resetting-induced localization produces the characteristic outer peaks, which are narrower than those in the diffusion-dominated regime due to the spatial constraints imposed by resetting.

\begin{figure}[!h]
    \centering
    \includegraphics[width=0.9\linewidth]{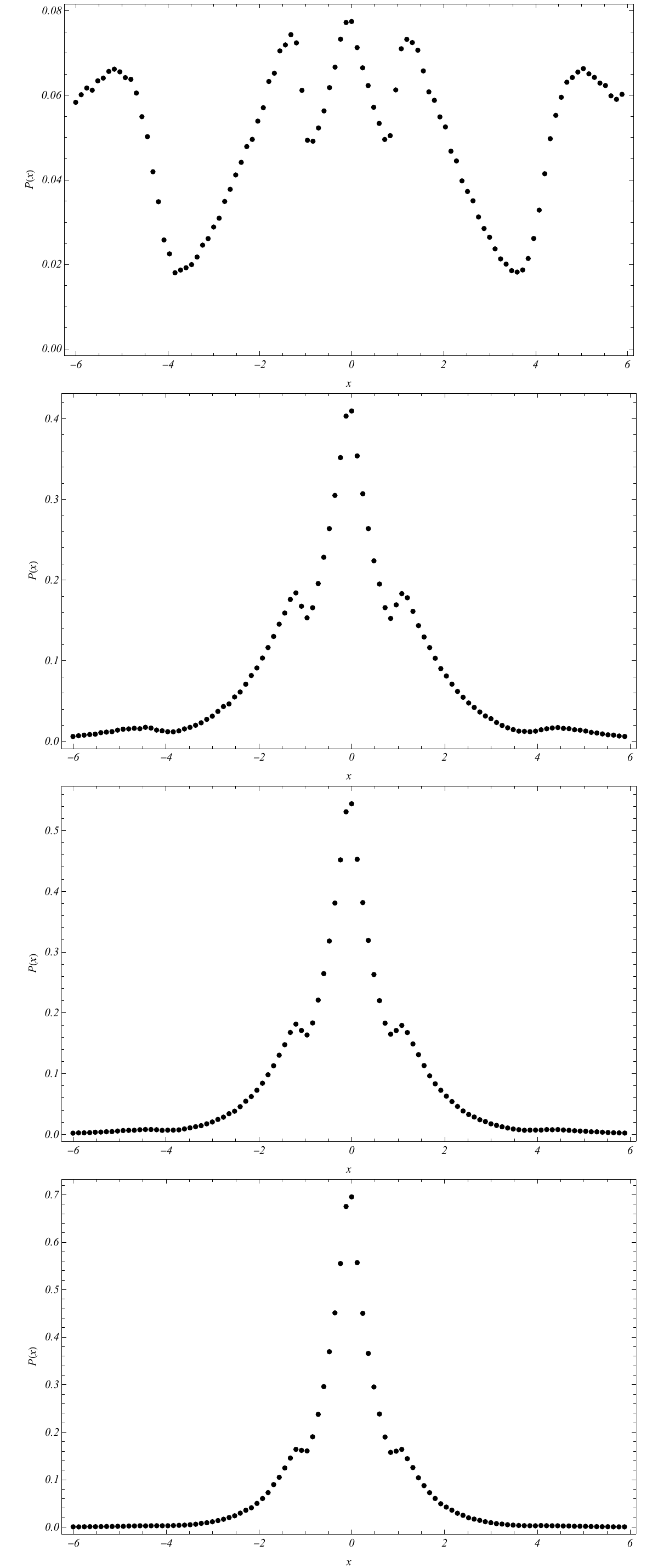}
    \caption{Stationary states for two level arctangent potential given by Eq.~\eqref{eq:multi-arctan-potential} for noise intensity $\sigma=1$ for various resetting rates, $r=0.1,1, 1.6, 2.5$ from top to bottom.}
    \label{fig:multi-arctan-states}
\end{figure}

Fially, Fig.~\ref{fig:multi-arctan-states} shows the stationary states for the potential defined in Eq.~\eqref{eq:multi-arctan-potential} with parameters $n=5$ and $\sigma=1$, shown for selected resetting rates $r$. 
As anticipated, the system develops statioary state with five distinct modes at small $r$ values. 
The modal structure follows the same general pattern observed for both the arctangent (Eq.~\eqref{eq:arctan-potential}) and Gaussian (Eq.~\eqref{eq:gaussian-potential}) potentials: the central peak originates from direct resetting effects, while the four outer peaks emerge from the interplay between resetting and the potential landscape. 
Specifically, the outer modes localize in regions of reduced potential steepness, with two additional maxima forming on the characteristic "shelves" created by local flattening of the potential. 
This shelf structure, unique to the potential give by Eq.~\eqref{eq:multi-arctan-potential}, modifies the spatial distribution of modes compared to the simpler potentials allowing for five-modal distribution of stationary states.

Following the same trend observed in trimodal-state potentials, increasing the resetting rate $r$ leads to growth of the central peak while suppressing the outer peaks. Crucially, this decay process is spatially non-uniform - outer modes located farther from the origin diminish more rapidly than their inner counterparts. 
This asymmetric decay creates a transitional regime where a trimodal state persists within a narrow range of resetting rates ($r \approx 1.6$-$2.6$ in our system). The emergence of this trimodal window stems from the reduced spatial exploration range caused by more frequent resetting events, which effectively curtails the particle's diffusion. 
While the exact boundaries of this transitional regime prove challenging to determine numerically due to gradual modal transitions, the system ultimately converges to a unimodal state at sufficiently high resetting rates.

Analysis for $\sigma=0.1$ (results not shown) revealed very similar behavior, with the difference that the range of $r$ for which the trimodal state occurs is significantly narrower, making its numerical analysis even more challenging. 
Additionally, the unimodal state is achieved at smaller values of $r$.

%%%%%%%%%%%%%%%%%%%%%%%%%%%%%%%%%%%%%%%%%%%%%%%%%%%%%%%%%%%%%%%%%%%%%%%%
%
%
% \clearpage
\section{Summary and conclusions \label{sec:summary}}
Resetting, a process that reinitializes the evolution of a system, can profoundly modify its stochastic dynamics, resulting in the emergence of non-equilibrium stationary states in unbounded potentials — states that would not typically exist without resetting. 
In this work, the stationary states for a random walk driven by Gaussian white noise were numerically analyzed for three exemplary potentials, each with a maximum at the origin and decreasing monotonically to $\pm \infty$. 
It was demonstrated that such potentials can support multimodal stationary states, specifically with $3$ and $5$ modes, where the modality depends on the noise intensity $\sigma$ and the resetting rate $r$.

For potentials capable of generating trimodal states, three distinct regimes were observed, depending on $\sigma$ and $r$: 
(i) a region of large $\sigma$, where bimodal states are observed; 
(ii) a regime of large $r$, where frequent resetting leads to unimodal states; and 
(iii) an intermediate region, where the interplay between resetting and diffusion in the potential creates trimodal states, forming a transition zone between the two extreme regimes.

For potentials that can exhibit five-modal states, the large $\sigma$ regime corresponds to states with the maximum number of modes. 
Additionally, an intermediate region emerges where trimodal states can be observed.

The first and most evident conclusion is that resetting can lead to the formation of stationary states with maxima that are neither located at the resetting point nor at the potential's minimum. 
Additional implications arise from the occurrence of five-modal states and the transition driven by changes in the resetting rate between three- and five-modal states. 
The presence of peaks on the `shelves', i.e., regions where the local slope of the potential is small, and the independence of these peaks from the existence of modes in regions farther from the origin, suggest that multimodal stationary states can also emerge in unbounded potentials that tend to $-\infty$ as $x \to \pm \infty$. 
This behavior is expected in potentials featuring such `shelves' and whose steepness still permits the existence of stationary states in the presence of resetting.

A further, more general conclusion is the similarity to other multimodal non-equilibrium stationary states with modes not located at the potential's minima, particularly those generated by L\'evy flights. 
In both cases, two key aspects governing the occurrence of stationary states can be identified. 
The first is the potential itself, which must exhibit an appropriate shape with a sufficient number of curvature maxima. 
The second is the presence of a process that, at repeated intervals, can relocate the particle to a position of higher energy (a higher value of the potential). 
For L\'evy flights, this process is facilitated by the heavy tails of the $\alpha$-stable noise distribution, which can place the particle at any point, while for the resetting processes studied here, it is the resetting mechanism itself, which typically returns the particle to the starting point. 
These similarities may provide deeper insights into the occurrence of multimodal non-equilibrium stationary states, extending beyond resetting processes.

%%%%%%%%%%%%%%%%%%%%%%%%%%%%%%%%%%%%%%%%%%%%%%%%%%%%%%%%%%%%%%%%%%%%%%%%
%
%
\section*{Acknowledgements}

This research received partial support from the funds assigned to AGH by Polish Ministry of Science and Higher Education. 
Authors gratefully acknowledge Polish high-performance computing infrastructure PLGrid (HPC Center: ACK Cyfronet AGH) for providing computer facilities and support within computational grant no. PLG/2025/018102

\section*{Data availability}
The data  (generated randomly using the model presented in the paper) that support the findings of this study are available from the corresponding author upon reasonable request.

%%%%%%%%%%%%%%%%%%%%%%%%%%%%%%%%%%%%%%%%%%%%%%%%%%%%%%%%%%%%%%%%%%%%%%%%
%
%

\section*{References}
%\bibliography{core-bibliography.bib}
\def\url#1{}

\end{document}